\documentclass[aps,prc,twocolumn,showpacs,superscriptaddress,floatfix]{revtex4}
\usepackage{graphicx}
\usepackage{dcolumn}
\usepackage{bm}
\usepackage{amsfonts}
\begin{document}

\title{Entrance Channel Dynamics of Hot and Cold Fusion Reactions\\Leading to Superheavy Elements}


\author{A.S. Umar}
\author{V.E. Oberacker}
\affiliation{Department of Physics and Astronomy, Vanderbilt University, Nashville, Tennessee 37235, USA}
\author{J.A. Maruhn}
\affiliation{Institut f\"ur Theoretische Physik, Goethe-Universit\"at, D-60438 Frankfurt am Main, Germany}
\author{P.-G. Reinhard}
\affiliation{Institut f\"ur Theoretische Physik, Universit\"at Erlangen, D-91054 Erlangen, Germany}

\date{\today}


\begin{abstract}
We  investigate the entrance channel dynamics for the reactions
$\mathrm{^{70}Zn}+\mathrm{^{208}Pb}$ and $\mathrm{^{48}Ca}+\mathrm{^{238}U}$
using the fully microscopic time-dependent Hartree-Fock (TDHF) theory coupled with a density constraint.
We calculate excitation energies and capture cross-sections relevant for the study of
superheavy formations. We discuss the deformation dependence of the ion-ion potential
for the $\mathrm{^{48}Ca}+\mathrm{^{238}U}$ system and perform an alignment angle
averaging for the calculation of the capture cross-section. The results show that this parameter-free approach 
can generate results in good agreement with experiment and other theories.
\end{abstract}
\maketitle

\section{Introduction}
One of the most fascinating research areas involving low-energy nuclear reactions
is the search for superheavy elements.
Experimentally, two approaches have been used for the synthesis of these elements,
one utilizing closed shell nuclei with lead-based targets (cold-fusion)~\cite{Ho00,Ho02}, the other
utilizing deformed actinide targets with $^{48}$Ca projectiles (hot-fusion)~\cite{Og04,Og07,Ho07}.
While both methods have been successful in synthesizing new elements the evaporation
residue cross-sections of the hot-fusion reactions were found to be as much as three
times larger than those of the cold fusion ones.
To pinpoint the root of this difference it is important to understand the details
of the entrance channel dynamics of these systems since the properties of the
dinuclear system at the capture point will strongly influence the outcome of the
reaction.
For light and medium mass systems the capture cross-section may be considered
to be the same as that for complete fusion, whereas for heavy systems leading to superheavy
formations the evaporation residue cross-section is dramatically reduced due to the
quasi-fission and fusion-fission processes, thus making the capture cross-section to
be essentially the sum of these two cross-sections.
What is also difficult to ascertain is the configuration of the composite system
namely, whether the system has a single-center compound-like configuration
or a dinuclear configuration accompanied by particle exchange.
Most dynamical models~\cite{Fa04,Fa05,AA03,NG09,AA09,FJ09} argue that for heavy systems a dinuclear
complex is formed initially and  the barrier structure and the excitation energy of this precompound
system will determine its survival to breaking up via  quasi-fission.
Furthermore, if the nucleus survives this initial state and evolves to a compound system
it can still fission due to its excitation.
Recent microscopic calculations have shown that the temperature (excitation energy) of the
compound systems strongly alters the barrier structures of potential energy surfaces,
with implications of very different fusion-fission cross-sections than those predicted
by zero temperature calculations~\cite{PN09}.

It is generally acknowledged that the TDHF theory provides a
useful foundation for a fully microscopic many-body theory of low-energy heavy-ion reactions
\cite{Ne82}. Recent three-dimensional TDHF calculations with no symmetry assumptions
and using modern Skyrme forces have been shown to accurately reproduce phenomena
determined by the early stages of the heavy-ion dynamics~\cite{GM08,UO09a,UOM08}.
Recently, we have developed the density-constrained TDHF (DC-TDHF) method~\cite{UO06b},
which is based on the generalization of the density constraint method developed
earlier~\cite{CR85}. We have shown that using the DC-TDHF method
ion-ion potential barriers can be accurately produced in most cases~\cite{UO06d,UO08a,UO09b},
as these calculations also depend on early stages of the ion-ion dynamics.
Furthermore, one-body energy
dissipation extracted from TDHF for low-energy fusion reactions was found to be in agreement with the
friction coefficients based on the linear response theory as well as those in models where
the dissipation was specifically adjusted to describe experiments~\cite{WL09}. All of these
new results suggest that TDHF dynamics may provide a good description of the early stages of
heavy-ion collisions.

Recently, we have also introduced an approach to extract excitation energies directly
from full microscopic TDHF calculations~\cite{UO10a}.
In this manuscript we perform TDHF calculations accompanied by density constraint calculations for
$\mathrm{^{70}Zn}+\mathrm{^{208}Pb}$ and
$\mathrm{^{48}Ca}+\mathrm{^{238}U}$ systems, which represent typical
examples of cold and hot fusion reactions leading to superheavy formation, respectively.
In addition to calculating the excitation energy at the capture point we also investigate
the capture cross-section and try to elucidate the differences between these two reactions.

\section{Theoretical Outline}
In this Section we will discuss issues pertaining to the application of TDHF to study
collisions  involving heavy reaction partners.
\subsection{TDHF Dynamics}
It is generally accepted that TDHF theory is a candidate for a microscopic theory
that may provide a unified approach to the description of diverse physical phenomena
such as fusion, deep-inelastic collisions, dinuclear and compound nucleus formation,
and possibly fission. Since TDHF is based on the independent-particle approximation
it can be interpreted as the semi-classical limit of a fully quantal theory thus allowing
a connection to macroscopic coordinates and providing insight about the collision process.
In this sense the TDHF dynamics can only compute
the semiclassical trajectories of the collective moments of the composite system as a
function of time. The presence of residual interaction, absent in TDHF, may produce
fluctuations and correlations which effect the mean values of these trajectories.

For TDHF collisions of light and medium mass systems as well as highly mass-asymmetric systems
fusion generally occurs immediately above the Coulomb barrier, while in heavier systems
there is an energy range above the barrier where fusion does not occur.
This phenomenon is the microscopic analogue of the macroscopic \textit{extra-push}
threshold~\cite{Sw82} and has been recently studied for TDHF collisions involving
heavy and nearly symmetric reaction partners~\cite{SA09}.
In the lower
part of this energy range deep-inelastic collisions are dominant while at slightly higher
energies the system develops a long-lived and pronounced neck reminiscent of a dinuclear
configuration. The outcome of these long-lived configurations is uncertain due to the
absence of quantal shell effects and fluctuations.
For these systems TDHF results in a compact configuration only for
energies considerably above the static potential energy surface.
However, despite the high energy, single-particle friction can quickly absorb this energy and
lead to a configuration that may be considered a thermal doorway state.
As long as the average single-particle excitation energy per nucleon in this doorway
state is less than the shell energy (about $4-8$~MeV) the details of the ground state
potential energy surface are still felt and shell correction energies influence the
TDHF dynamics.
It is precisely for this reason that the DC-TDHF approach allows us to
reproduce ion-ion interaction barriers for heavy-ion collisions.

\subsection{DC-TDHF and Excitation Energy}
In the DC-TDHF approach~\cite{UO06b}
the TDHF time-evolution takes place with no restrictions.
At certain times during the evolution the instantaneous density is used to
perform a static Hartree-Fock minimization while holding the neutron and proton densities constrained
to be the corresponding instantaneous TDHF densities. In essence, this provides us with the
TDHF dynamical path in relation to the multi-dimensional static energy surface
of the combined nuclear system. The advantages of this method in comparison to other mean-field
based microscopic methods such as the constrained Hartree-Fock (CHF) method are obvious. First,
there is no need to introduce artificial constraining operators which assume that the collective
motion is confined to the constrained phase space: second, the static adiabatic approximation is
replaced by the dynamical analogue where the most energetically favorable state is obtained
by including sudden rearrangements and the dynamical system does not have to move along the
valley of the potential energy surface. In short we have a self-organizing system which selects
its evolutionary path by itself following the microscopic dynamics.
All of the dynamical features included in TDHF are naturally included in the DC-TDHF calculations.
These effects include neck formation, mass exchange,
internal excitations, deformation effects to all order, as well as the effect of nuclear alignment
for deformed systems.
In the DC-TDHF method the ion-ion interaction potential is given by
\begin{equation}
V(R)=E_{\mathrm{DC}}(R)-E_{\mathrm{A_{1}}}-E_{\mathrm{A_{2}}}\;,
\label{eq:vr}
\end{equation}
where $E_{\mathrm{DC}}$ is the density-constrained energy at the instantaneous
separation $R(t)$, while $E_{\mathrm{A_{1}}}$ and $E_{\mathrm{A_{2}}}$ are the binding energies of
the two nuclei obtained with the same effective interaction.
In writing Eq.~(\ref{eq:vr}) we have introduced the concept of an adiabatic reference state for
a given TDHF state. The difference between these two energies represents the internal energy.
The adiabatic reference state is the one obtained via the density constraint calculation, which
is the Slater determinant with lowest energy for the given density with vanishing current
and approximates the collective potential energy~\cite{CR85}.
We would like to
emphasize again that this procedure does not affect the TDHF time-evolution and
contains no free parameters or normalization.

Ion-ion interaction potentials calculated using DC-TDHF correspond to the
configuration attained during a particular TDHF collision. For light and
medium mass systems as well as heavier systems for which fusion is
the dominant reaction product, DC-TDHF
gives the fusion barrier with an appreciable but relatively small energy dependence.
On the other hand, for reactions leading to superheavy systems fusion is
not the dominant channel at barrier top energies. Instead the system sticks
in some dinuclear configuration with possible break-up after exchanging a
few nucleons. The long-time evolution to break-up is beyond the scope of
TDHF due to the absence of quantal shell effects and fluctuations.
As we increase the energy above the barrier this phenomenon gradually
changes to the formation of a truly composite object. This is somewhat
similar to the extra-push phenomenon discussed in phenomenological
models.
For this reason the energy dependence of the DC-TDHF interaction barriers
for these systems is not just due to the dynamical effects for the same final
configuration but actually represent different final configurations.

The calculation of the excitation energy is achieved by dividing the TDHF
motion into a collective and intrinsic part~\cite{UO10a}. The major assumption in
achieving this goal is to assume that the collective part is primarily determined by
the density $\rho(\mathbf{r},t)$ and the current $\mathbf{j}(\mathbf{r},t)$. Consequently,
the excitation energy can be formally written as
\begin{equation}
E^{*}(t)=E_{TDHF}-E_{coll}\left(\rho(t),\mathbf{j}(t)\right)\;,
\label{eq:ex}
\end{equation}
where $E_\mathrm{TDHF}$ is the total energy of the dynamical system, which is a conserved quantity,
and $E_\mathrm{coll}$ represents the
collective energy of the system. In the next step we break up the collective energy into two parts
\begin{equation}
E_{coll}\left(t\right)= E_{kin}\left(\rho(t),\mathbf{j}(t)\right) + E_{DC}\left(\rho(t)\right)\;,
\end{equation}
where $E_\mathrm{kin}$ represents the kinetic part and is given by
\begin{equation}
E_{kin}\left(\rho(t),\mathbf{j}(t)\right)=\frac{m}{2}\int\;\textrm{d}^{3}r\;\mathbf{j}^2(t)/\rho(t)\;,
\end{equation}
which is asymptotically equivalent to the kinetic energy of the
relative motion, $\frac{1}{2}\mu\dot{R}^2$, where $\mu$ is the
reduced mass and $R(t)$ is the ion-ion separation distance.
The dynamics of the ion-ion separation
$R(t)$ is provided by an unrestricted TDHF run thus allowing us to deduce
the excitation energy as a function of the distance parameter, $E^{*}(R)$.

\section{Results}
Calculations were done in 3-D geometry and using the full Skyrme force (SLy4)~\cite{CB98}
without the center-of-mass correction as described in Ref.~\cite{UO06}.
We have performed density constraint calculations every $20$ time steps.
For the calculation of the ion-ion separation distance $R$ we use the hybrid method, which
relates the coordinate to the quadrupole moment for small $R$ values, as described in
Ref.~\cite{UO09b}. The accuracy of the density constraint calculations is
commensurate with the accuracy of the static calculations.

\subsection{$^{48}$Ca+$^{238}$U System}
\begin{figure}[!htb]
\includegraphics*[scale=0.35]{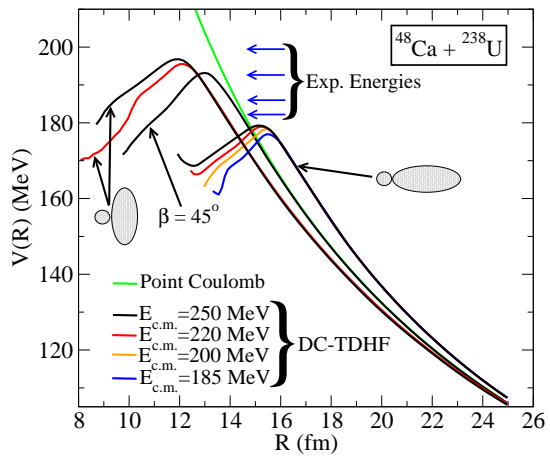}
\caption{\label{fig1} (color online)
Potential barriers, $V(R)$,  for the $^{48}$Ca+$^{238}$U system obtained from
DC-TDHF calculations using Eq.~\protect(\ref{eq:vr}) as a function of $E_\mathrm{c.m.}$ energy and
for selected orientation angles $\beta$ of the $^{238}$U nucleus. Also, shown are the experimental c.m. energies.}
\end{figure}

As an example of superheavy formation from a hot-fusion reaction we have studied the $^{48}$Ca+$^{238}$U system.
Hartree-Fock (HF) calculations produce a  spherical $^{48}$Ca nucleus, whereas $^{238}$U has a large axial deformation.
The large deformation of $^{238}$U is expected to strongly influence the interaction barriers for this system.
This is shown in Fig.~\ref{fig1}, which shows the interaction barriers, $V(R)$,
calculated using the DC-TDHF method as a function of c.m. energy and for three different
orientations of the $^{238}$U nucleus. The alignment angle $\beta$ is the angle
between the symmetry axis of the $^{238}$U nucleus and the collision axis.
Also shown in Fig.~\ref{fig1} is the point Coulomb potential corresponding to this collision.
The deviations from the point Coulomb potential at large $R$ values are due to the deformation
of the $^{238}$U nucleus.
We first notice that the barriers corresponding to the polar orientation ($\beta=0^{o}$) of the
$^{238}$U nucleus are much lower and peak at larger ion-ion separation distance $R$.
On the other hand, the barriers corresponding to the equatorial orientation of $^{238}$U ($\beta=90^{o}$) are
much higher and peak at smaller $R$ values. For the intermediate values of $\beta$ the barriers
rise rapidly as we increase the orientation angle from $\beta=0^{o}$, as can be seen for $\beta=45^{o}$.
The rise in the barrier height as a function of increasing $\beta$ values is not linear but seems to
rise more rapidly for smaller $\beta$ values.
We also see that for lower energies central collisions with polar orientation of $^{238}$U are the only orientations which result in
the sticking of the two nuclei, while the equatorial orientations of $^{238}$U result in a deep-inelastic
collision.
Also, shown in Fig.~\ref{fig1} are the experimental energies~\cite{Ho07,Og07} for this reaction.
We observe that all of the experimental energies are above the barriers obtained for the
polar alignment of the $^{238}$U nucleus.

Furthermore, the potentials shown in Fig.~\ref{fig1} display a very strong energy dependence.
Detailed analysis of the TDHF time-evolution and density profiles show that at lower c.m. energies
and for central collisions the polar
configuration of $^{238}$U leads to a dinuclear system where both nuclei maintain their
cores and exchange nucleons. Non-central collisions at these energies result in deep-inelastic
fragments. As we mentioned above, for these low energy collisions the equatorial collisions
result in deep-inelastic reaction products even for central collisions.
At higher energies the system forms a true composite with overlapping
cores. These are the potentials which should be used for calculating capture cross-sections.

For the formation of a superheavy system the excitation energy carries great significance
since a high-excitation energy at the capture point would result in quasi-fission events
while high excitation of the compound nucleus will lead to fusion-fission.
Many factors play a role in building up the excitation energy. As we have discussed in the
Introduction, recent TDHF calculations suggest that during the early phase of the ion-ion
collisions TDHF theory may provide a good approximation for the transfer of the initial
kinetic energy into internal and collective excitations via the dynamical evolution of shell effects.
Naturally, the excitation also depends on other nuclear properties such as deformation and alignment,
mass asymmetry in the entrance channel, and impact parameter.
\begin{figure}[!htb]
\includegraphics*[scale=0.35]{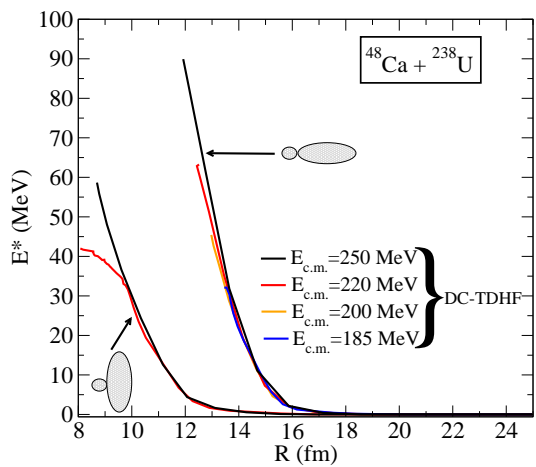}
\caption{\label{fig2} (color online)
Excitation energy, $E^{*}(R)$, in a central collision for various
$E_\mathrm{c.m.}$ energies and for alignment angles angles $\beta=0^{o}$ and $\beta=90^{o}$ of the $^{238}$U nucleus.}
\end{figure}
In Fig.~\ref{fig2} we show the excitation energy $E^{*}(R)$ as a function of c.m. energy
and for two alignment angles ($\beta=0^{o}$ and $\beta=90^{o}$ ) of the $^{238}$U nucleus.
The excitation energy curves start at zero excitation when the two nuclei are far apart,
which also provides a test for the numerical accuracy of the calculation.
We note that the system is
excited much earlier during the collision process for the polar alignment of the $^{238}$U nucleus
and has a higher excitation than the corresponding collision for the equatorial orientation.
Only two curves are shown for the equatorial collision since at lower energies we have
deep-inelastic collisions for this alignment.
\begin{figure}[!htb]
\includegraphics*[scale=0.35]{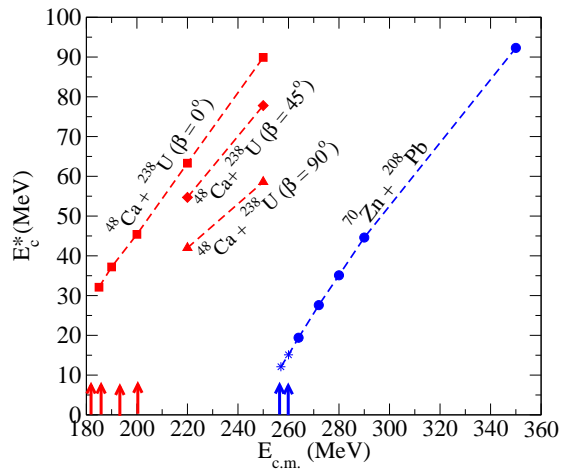}
\caption{\label{fig3} (color online)
Excitation energy in a central collision at the potential minimum inside the barrier, $E_{c}^{*}$, as a function of $E_\mathrm{c.m.}$ for the $^{48}$Ca+$^{238}$U
and the $^{70}$Zn+$^{208}$Pb systems.
For the case of $^{48}$Ca+$^{238}$U system $^{238}$U alignment angles of $\beta=0^{o},45^{o}$,  and $\beta=90^{o}$
are shown.}
\end{figure}
We note that the highest point reached for these excitation curves is chosen
to be the point where the nuclei almost come to a stop inside the barrier, which corresponds to a nearly zero collective
kinetic energy. For energies for which the collision outcome is capture
this would be the excitation energy at the capture point.

In Fig.~\ref{fig3} we show the excitation energy as a function of the center-of-mass energy for both
$^{48}$Ca+$^{238}$U and $^{70}$Zn+$^{208}$Pb systems calculated at the potential minimum inside
the barrier.
The excitation energy curves increase with a typical  linear slope as a function of $E_\mathrm{c.m.}$. For the case of
$^{48}$Ca+$^{238}$U system we show the excitation energies for  $^{238}$U alignment angles of $\beta=0^{o},45^{o}$,
and $\beta=90^{o}$.
The upward arrows in the bottom of the figure denote the experimental energies for these reactions.
For larger alignment angles of $^{238}$U, the curves start at a higher center-of-mass energy since at lower energies
we only have deep-inelastic collisions. The lowest two points for the $^{70}$Zn+$^{208}$Pb curve (shown as stars) are
simply the extrapolated values down to the experimental energies.

While at first glance the excitation energies for the $^{48}$Ca+$^{238}$U system look much higher than those
for the $^{70}$Zn+$^{208}$Pb system, this may be somewhat misleading. The reason is that the excitations
for the $^{48}$Ca+$^{238}$U system must be angle averaged for different alignments of the $^{238}$U nucleus
and also weighted by the alignment probability discussed below. So, for example the $\beta=0^{o}$ excitation
at a particular center-of-mass energy will be multiplied by the $\sin(\beta)$ factor in the integration weight and
will make a zero contribution. The value of $\sin(\beta)$ increases with $\beta$, but at the
same time the excitation energy decreases.
Since TDHF calculations at the experimental energies only yield dinuclear configurations for the small
values of $\beta$ but rapidly move into the deep-inelastic domain for larger values we could not compute
this average in practice.
Naturally, a fully quantal system will have a certain probability for resulting in a dinuclear configuration
for all values of $\sin(\beta)$ at these energies and the excitation energy will be smaller than the one
one shown for the $\beta=0^{o}$ case.

As we have discussed earlier, for systems leading to superheavy formation the evaporation residue cross-section
is customarily represented in terms of the various phases of the reaction process as
\begin{equation}
\sigma_{\rm ER}=\sigma_{\rm capture}\cdot P_{\rm CN}\cdot P_{\rm survival}\;,
\end{equation}
where $\sigma_{\rm ER}$ denotes the evaporation residue cross-section for the superheavy system,
$\sigma_{\rm capture}$ is the capture cross-section for the two-ions, $P_{\rm CN}$ is the probability of forming
a compound nucleus, and $P_{\rm survival}$ is the probability that this compound system survives
various breakup and fission events. The calculations presented here can only address the
capture cross-section for these systems since the subsequent reaction possibilities are beyond
the scope of the TDHF theory. For most light systems for which fusion is the dominant reaction
product $\sigma_{\rm capture}$ and $\sigma_{\rm ER}$ are essentially the same and equal to the fusion
cross-section, $\sigma_{\rm fusion}$. Instead, for reactions involving superheavy formations we have
\begin{equation}
\sigma_{\rm \rm capture}=\sigma_{\rm QF}+\sigma_{\rm FF}+\sigma_{\rm ER}\;,
\end{equation}
where $\sigma_{\rm QF}$ and $\sigma_{\rm FF}$ denote the quasi-fission and fusion-fission cross-sections, respectively.
For these reactions the evaporation residue cross-section, $\sigma_{\rm ER}$, is very small and therefor the
capture cross-section is to a large extent equal to sum of the two fission cross-sections.
Furthermore, the distinction between deep-inelastic reactions  and quasi-fission is somewhat difficult and usually
achieved by setting windows for fragment masses of $A_{f} = A_{\rm CN}/2 \pm 20$ and on their kinetic energy.
\begin{figure}[!htb]
\includegraphics*[scale=0.35]{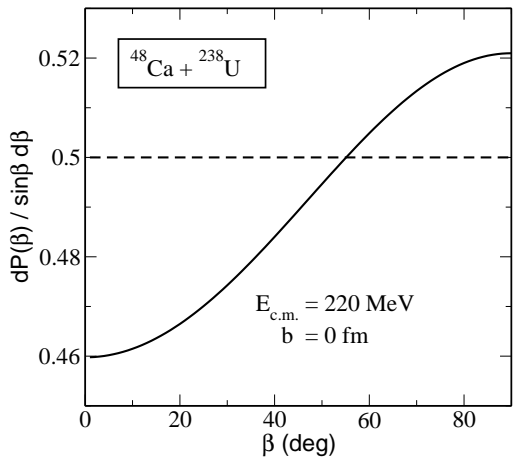}
\caption{\label{fig4} Dynamic alignment due to Coulomb
excitation of $^{238}$U. Shown is the orientation probability
as a function of the Euler angle $\beta$ in a central collision at internuclear
distances $R=1500$~fm (dashed horizontal curve) and at $R=20$~fm at
center-of-mass energy of $E_\mathrm{c.m.}=220$~MeV (solid curve).
}
\end{figure}

For the calculation of the capture cross-section we need to average over all possible alignments of
the $^{238}$U nucleus~\cite{DH00,GH00,DN02,UO06e,GM09} as given by
\begin{equation}
\label{eq:scapture}
\sigma_{\rm capture}(E_{\mathrm{c.m.}}) = \int_{0}^{\pi} \mathrm{d}\beta\;\sin(\beta)\; \frac{\mathrm{d}P}{\mathrm{d}\beta\sin(\beta)}\; 
\sigma(E_{\mathrm{c.m.}},\beta)\;,
\end{equation}
where $\mathrm{d}P/(\mathrm{d}\beta\;\sin(\beta))$ represents the alignment probability and $\sigma(E_{\mathrm{c.m.}},\beta)$
is the capture cross-section associated with a particular alignment. In practice, we calculate the
alignment probability due to Coulomb excitation as described in Ref.~\cite{UO06e}.
Due to the relatively small charge of the $^{48}$Ca nucleus this probability is in the range $0.46-0.52$
as shown in Fig.~\ref{fig4} and does not vary appreciably with energy.
\begin{figure}[!htb]
\includegraphics*[scale=0.35]{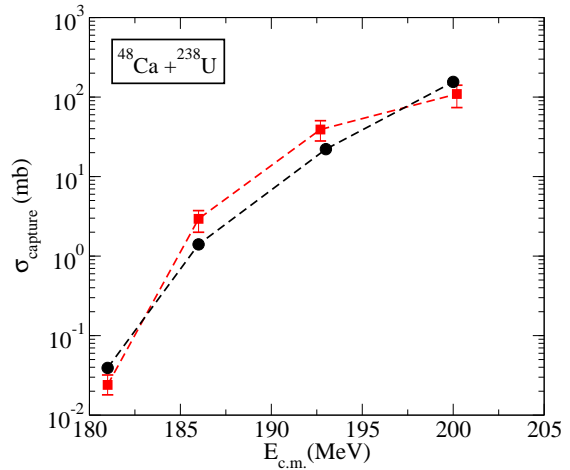}
\caption{\label{fig5} (color online) Capture cross-sections for the $^{48}$Ca+$^{238}$U system
as a function of $E_\mathrm{c.m.}$ energy (black circles). Also, shown are the experimental cross-sections
(red squares)~\cite{Og07}.}
\end{figure}
One important fact to notice in the
cross-section formula given in Eq.~(\ref{eq:scapture}) is that 
the cross-section is multiplied by the $\sin(\beta)$ factor, which renders the contribution
originating from the lowest barriers at small values of $\beta$ to be very small.
However, unlike the calculation of excitation energy, which requires a TDHF collision at exactly the
same experimental energy, the potential barriers are obtained by performing TDHF calculations
at higher energies and using the density constraint to calculate the ion-ion potentials.
For a consistent calculation of fusion cross sections above
and below the barrier energies we have adopted the commonly used
Incoming Wave Boundary Condition (IWBC) method~\cite{Raw64,LP84}.
In practice, we have varied the alignment angle $\beta$ in $10^{o}$ steps between $0^{o}$ and $90^{o}$.
In Fig.~\ref{fig5} we show the capture cross-sections for the $^{48}$Ca+$^{238}$U system
as a function of $E_\mathrm{c.m.}$ energy (black circles). Also, shown are the experimental cross-sections
(red squares)~\cite{Og07}.
While the agreement with experiment is not perfect it is still remarkably good since these calculations
contain no free parameters.

\subsection{$^{70}$Zn+$^{208}$Pb System}
\begin{figure}[!htb]
\includegraphics*[scale=0.35]{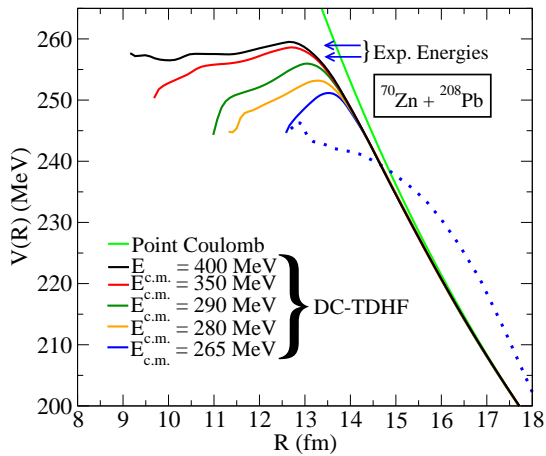}
\caption{\label{fig6} (color online)
Potential barriers, $V(R)$,  for the $^{70}$Zn+$^{208}$Pb system obtained from
DC-TDHF calculations using Eq.~\protect(\ref{eq:vr}), for various $E_\mathrm{c.m.}$ energies as indicated.
}
\end{figure}

We have also investigated the cold-fusion reaction of $^{70}$Zn+$^{208}$Pb, which
lead to the same superheavy element $Z=112$ (but a different isotope). The Hartree-Fock calculations for $^{70}$Zn
and $^{208}$Pb produce a spherical nucleus in both cases.
In Fig.~\ref{fig6} we plot the potential barriers for the $^{70}$Zn+$^{208}$Pb system obtained from
DC-TDHF calculations using Eq.~\protect(\ref{eq:vr}) as a function of $E_\mathrm{c.m.}$ energy.
Also shown is the point Coulomb potential for two spherical nuclei. Unlike light nuclei, the
deviations from the Coulomb trajectory start relatively early ($15-16$~fm) due to the large charge
of the two ions.
A similar behavior to the $^{48}$Ca+$^{238}$U system is observed for the $^{70}$Zn+$^{208}$Pb system, namely at lower energies
nuclear densities show a dinuclear character for central collisions and result in deep-inelastic
collision at non-central impact parameters. The lowest barrier in Fig.~\ref{fig6} shows the case
at $E_\mathrm{c.m.}=265$~MeV where even for a central collision the result is deep-inelastic
as can be seen from the dotted blue curve. Only at the highest two energies the densities
corresponding to central collisions show a more composite character, which may be identified as capture,
and result in a similar density distribution for non-central collisions as well.
We also observe the flattening of the ion-ion potentials as the nuclear overlap increases
at smaller $R$ values, which seems to be a general behavior for collisions of two heavy
nuclei.
\begin{figure}[!htb]
\includegraphics*[scale=0.35]{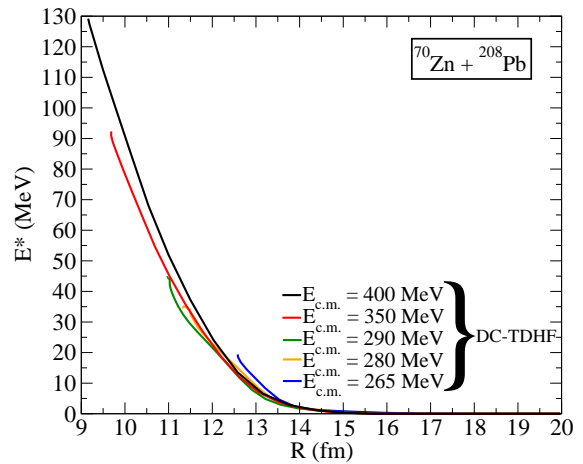}
\caption{\label{fig7} (color online)
Excitation energy, $E^{*}(R)$ for the $^{70}$Zn+$^{208}$Pb system and for various $E_\mathrm{c.m.}$ energies as indicated.}
\end{figure}
In Fig.~\ref{fig7} we show the excitation energy, $E^{*}(R)$, as a function of the $E_\mathrm{c.m.}$
energy for the $^{70}$Zn+$^{208}$Pb system. Again, the highest value attained is at the point
where the two nuclei come to a stop inside the ion-ion potential. These energies were shown
in Fig.~\ref{fig3} as a function of $E_\mathrm{c.m.}$ energy.
 
\begin{figure}[!htb]
\includegraphics*[scale=0.35]{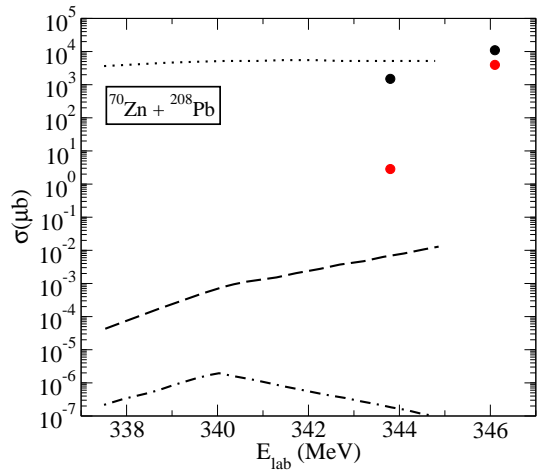}
\caption{\label{fig8} (color online)
Cross-sections (in $\mu$b) from Ref.~\protect\cite{GH00} as a function of $E_\mathrm{lab}$ energy for the $^{70}$Zn+$^{208}$Pb system.
The dotted curve shows the quasi-fission, the dashed curve  the fusion, and the dot-dashed curve the $1n$ evaporation
residue cross-section. Our calculations at the two experimental energies corresponding to the two potential energies leading
to capture in Fig.~\ref{fig6} are shown as filled circles.}
\end{figure}

We have also calculated the capture cross-section for the $^{70}$Zn+$^{208}$Pb system.
Since we are dealing with two spherical nuclei no angle averaging is necessary for this system.
However, we could not find experimental measurements for the capture cross-section.
For comparison we have used a model calculation where the cross-sections for the synthesis of
superheavy elements were analyzed using the concept of a dinuclear system~\cite{GH00}.
The authors calculated capture or quasi-fission, fusion, and $1n$ evaporation residue cross-sections
which reproduced a single $1n$ evaporation residue experimental cross-section.
For comparison we have plotted our capture cross-sections together with their findings for the
two experimental laboratory energies~\cite{Ho02} of $E_{\mathrm{lab}}=343.8$~MeV and $346.1$~MeV
in Fig.~\ref{fig8}.
Our calculations at the two experimental energies corresponding to the two potential energy curves leading
to capture in Fig.~\ref{fig6} are shown as filled circles.
Differences in laboratory energies between our results and the results of the model calculation~\cite{GH00}
stem from the fact that model calculations make use of the compound nucleus excitation energy
and relate this to the laboratory energy. We have directly used the experimental energies from Ref.~\cite{Ho02}
since we do not compute the compound nucleus excitation energy.
On the other hand, the dependence of the quasi-fission cross-section on laboratory energy is very flat
as shown in Fig.~\ref{fig8}.
As we see the calculated cross-sections at the higher laboratory energy of $346.1$~MeV is in a reasonable
agreement with the model calculation for both potential energy curves, whereas for the lower laboratory
energy of $343.8$~MeV  the lower potential energy curve substantially underestimates the model calculation.
We may conclude from this observation that the higher potential energy curve provides a better representation
of the inner part of the ion-ion barrier. However, since we cannot rule out the lower potential energy curve
as a capture event based on our TDHF results we have shown both results.

\section{Summary}
In this work we have investigated two systems which are known to produce the element $Z=112$ in experiments.
We find that the collisions of such heavy systems have very different characteristics than
the TDHF calculations of light and medium-mass systems.
We have used the density constraint along with TDHF to obtain ion-ion interaction potentials
and excitation energies.
The dependence of the ion-ion potential on the deformation of the $^{238}$U nucleus
was studied.
The calculated capture cross-sections are found to be in reasonable agreement with data
and other model calculations.

The fully microscopic TDHF theory has shown itself to be rich in
nuclear phenomena and continues to stimulate our understanding of nuclear dynamics.
The time-dependent mean-field studies seem to show that the dynamic evolution
builds up correlations that are not present in the static theory.
While modern Skyrme forces provide a much better description of static nuclear properties
in comparison to the earlier parametrizations there is a need to obtain even better
parametrizations that incorporate deformation and scattering data into the fit process.

\begin{acknowledgments}
This work has been supported by the U.S. Department of Energy under grant No.
DE-FG02-96ER40963 with Vanderbilt University, and by the German BMBF
under contract Nos. 06F131 and 06ER142D. One of us (A.S.U.) thanks for the support by the Hessian
LOEWE initiative through the Helmholtz International Center for FAIR (HICforFAIR) during
his stay in Frankfurt.
\end{acknowledgments}

\end{document}